\documentclass[iop]{emulateapj}
\usepackage{wasysym}
\usepackage[autostyle, english = american]{csquotes} 
\usepackage{soul}
\usepackage{natbib}
\usepackage[dvipsnames]{xcolor}
\usepackage{url}

\MakeOuterQuote{"}

\def\candidate{KOI-368.01}
\def\star{KOI-368}

\shortauthors{Ahlers et al.}

\begin{document}

\title{Spin-Orbit Alignment for 110-Day-Period KOI368.01 from Gravity Darkening}

\author{John P. Ahlers\footnotemark[*], Shayne A. Seubert\footnotemark[*], Jason W. Barnes} 
\affil{Department of Physics} 
\affil{University of Idaho}
\affil{875 Perimeter Dr. Stop 440903}
\affil{Moscow, ID 83844-0903, USA}  

\newpage
\footnotetext[*]{These authors contributed equally to this work}
\footnotetext[1]{A complete list of spin-orbit misaligned planets is available at http://www.physics.mcmaster.ca/\texttt{\char`\~}rheller/}
\begin{abstract} 
We fit the \emph{Kepler} photometric light curve of the KOI-368 system using an oblate, gravity-darkened stellar model in order to constrain its spin-orbit alignment. We find that the system is relatively well-aligned with a sky-projected spin-orbit alignment of $\lambda=10^{\circ}\pm2^{\circ}$, a stellar obliquity of $\psi=3^{\circ}\pm7^{\circ}$, and a true spin-orbit alignment of $\varphi=11^{\circ}\pm3^{\circ}$. Although our measurement differs significantly from zero, the low value for $\varphi$ is consistent with spin-orbit alignment. We also measure various transit parameters of the KOI-368 system: $R_{\star}=2.28\pm0.02 R_{\odot}$, $R_{p}=1.83\pm0.02R_{jup}$, and $i=89.221^{\circ}\pm0.013^{\circ}$. This work shows that our gravity-darkened model can constrain long-period, well-aligned planets and M-class stars orbiting fast-rotators, allowing for measurement of a new subcategory of transiting bodies.
\end{abstract}

 
\section{\textbf{INTRODUCTION}}

Main-sequence stars earlier than spectral type $\sim$F6 are expected to rotate rapidly due to their radiative exteriors \citep{2009ApJ...705..683B}. This induces the stellar figure to become oblate, which causes the star's photosphere to be up to several thousand Kelvin hotter at the poles than at the equator, leading to higher polar luminosity. This effect, called gravity darkening, was first predicted by \citet{1924MNRAS..84..665V}. Gravity darkening causes asymmetric light curves for misaligned transiting candidates \citep{2009ApJ...705..683B}, and has been used to constrain spin-orbit alignments for significantly misaligned candidates \citep{2011ApJS..197...10B}. This work will show that this method can also constrain spin-orbit aligned systems with relatively symmetric transit light curves for eclipsing objects.

The measurement of the angle between the inclination of a planet's orbit normal and parent star's spin axis, spin-orbit alignment ($\varphi$), can tell us more about the formation and evolution of that system. Evidence shows that a wide variety of planetary system types exist, including many short and long period spin-orbit misaligned planets \citep{10.1086/659427, 2010MNRAS.402L...1P, 2009A&A...506..377T, 2012A&A...537A.136G}\footnotemark[1]. We can use constrained spin-orbit alignments to compare planetary formation of extrasolar planets to that in our own planetary system. We propose an improved method for finding the spin-orbit alignment based off of \citet{2011ApJS..197...10B}, allowing for constraint of previously unmeasurable systems.

There are several existing methods for calculating the stellar obliquity and the sky-projected spin-orbit alignment, including the Rossiter-McLaughlin effect, stroboscopic starspots, Doppler tomography, asteroseismic determination of obliquity and gravity darkening. The Rossiter-McLaughlin technique uses Doppler shifts in radial velocity measurements during the eclipse of the primary star. \citep{1924ApJ....60...15R, 1924ApJ....60...22M}. Stroboscopic starstpots can be used to constrain an aligned system because the planet will cross the same starspot each time the planet transits. However, if the system is misaligned, the planet will cross the starspot very infrequently \citep{2011ApJS..197...14D}, and the method will fall short. Doppler tomography is able to achieve similar results to those possible from gravity darkening, but this method needs high signal-to-noise radial velocity follow up measurements \citep{2012A&A...543L...5G}. Asteroseismic determination of obliquity can constrain the obliquity, but not the projected spin-orbit alignment \citep{2013arXiv1302.3728C}, so other measurements are required to find the spin-orbit alignment.   

Gravity darkening constrains {\em both} the stellar obliquity and the sky-projected spin-orbit alignment simultaneously. \citet{2011ApJS..197...10B} used gravity darkening to establish the spin-orbit misalignment in Kepler Object of Interest number 13 (KOI-13) system. \citet{2012ApJ...756...66H} state that "For such rapid rotators, asymmetries in the transit light curve may be used to determine the parameters only if the spin-orbit angle is large;" however, we show here that we can constrain such rapid rotating systems, even if the spin-orbit angle is small.

In this paper, we show that our gravity-darkened model can constrain long-period, well-aligned planets and M-class stars orbiting fast-rotators, allowing for measurement of a new subcategory of transiting bodies. In Section 2, we describe our steps for data collection and preparation. In Section 3, we outline the gravity-darkened model that we use to fit the \emph{Kepler} transit curve, and in Section 4 we list our constrained parameters for the KOI-368 system. We discuss implications of this work in Section 5. In section 6 we compare our results to \citet{arXiv:1307.2249v3}. This work can be applied to the formation and evolution of intermediate-period planets orbiting fast-rotating stars and eclipsing binary systems; however, actual application of these concepts is beyond the scope of this paper.  

\section{Preparing the KOI-368.01 Light Curve} \label{section:observations}

\citet{2011ApJ...736...19B} first noted \candidate~as a transiting planet candidate. This KOI is particularly interesting because of the parent star, \star's, early spectral type ($T_\mathrm{eff}=9257$~K); because the star's brightness ($m_\mathrm{Kepler}=11.375$) leads to a high total signal-to-noise ratio (SNR) for the transit (SNR=1866.4); and because of the relatively long orbital period for its companion of 110 days. We summarize the original star and planet candidate parameters in Table \ref{table:parameters}.

More recently, follow-up spectroscopy of \star~from the \emph{Kepler} Community Follow-up Observation Program (CFOP) showed a high degree of rotational broadening of the stellar absorption lines. That broadening allows measurement of the star's projected rotational velocity, conventionally denoted $v\sin(i)$ ($v\cos(\psi)$ using our parameter definitions, where $\psi$ is the star's obliquity relative to the plane of the sky). The CFOP-measured $v\sin(i)$ value is 90~km/s, as measured by the TRES Echelle Spectrograph of the Smithsonian Astrophysical Observatory. Hence, \star~is a rapid rotator, and may therefore show sufficient gravity darkening to allow us to measure the relative alignment angle $\varphi$ between the stellar rotation pole and the orbit normal.

To do so, we first acquire SAP\_FLUX photometric timeseries for \star~from the Mikulski Archive for Space Telescopes (MAST) \emph{Kepler} database, including both short cadence (60 second integration time) and long cadence (30 minute integration time) photometry. We use \emph{Kepler} Quarter 0 (Q0) through Quarter 16 (Q16) public data, totalling 67601 short cadence and 60491 long cadence data points. During Q0-Q7 and Q10-Q16 the spacecraft only used long cadence for \star, but during Q8 and Q9 it used short and long cadence mode. When using both the long cadence and available short cadence data from Q0-Q16 we keep track of the specific integration time used for each data point.

Transiting planet candidate \candidate~shows a total of 11 long cadence and 2 short cadence transits within the 17 quarters. One transit is missing because it is in a data gap. This missing transit is the eighth in the sequence.

We performed several steps to prepare both the short and long cadence data for fitting. We first normalize each quarter's flux by dividing each point by its quarter's median value. We then glue all short cadence quarters together and all long cadence quarters together, and use a median boxcar filter with a period of triple the transit duration (42 hours) to correct for long-term instrument response variations. We show the full processed Q0-Q16 timeseries in Figure 1. 

\begin{figure}[tbhp]
\epsscale{1}
\plotone{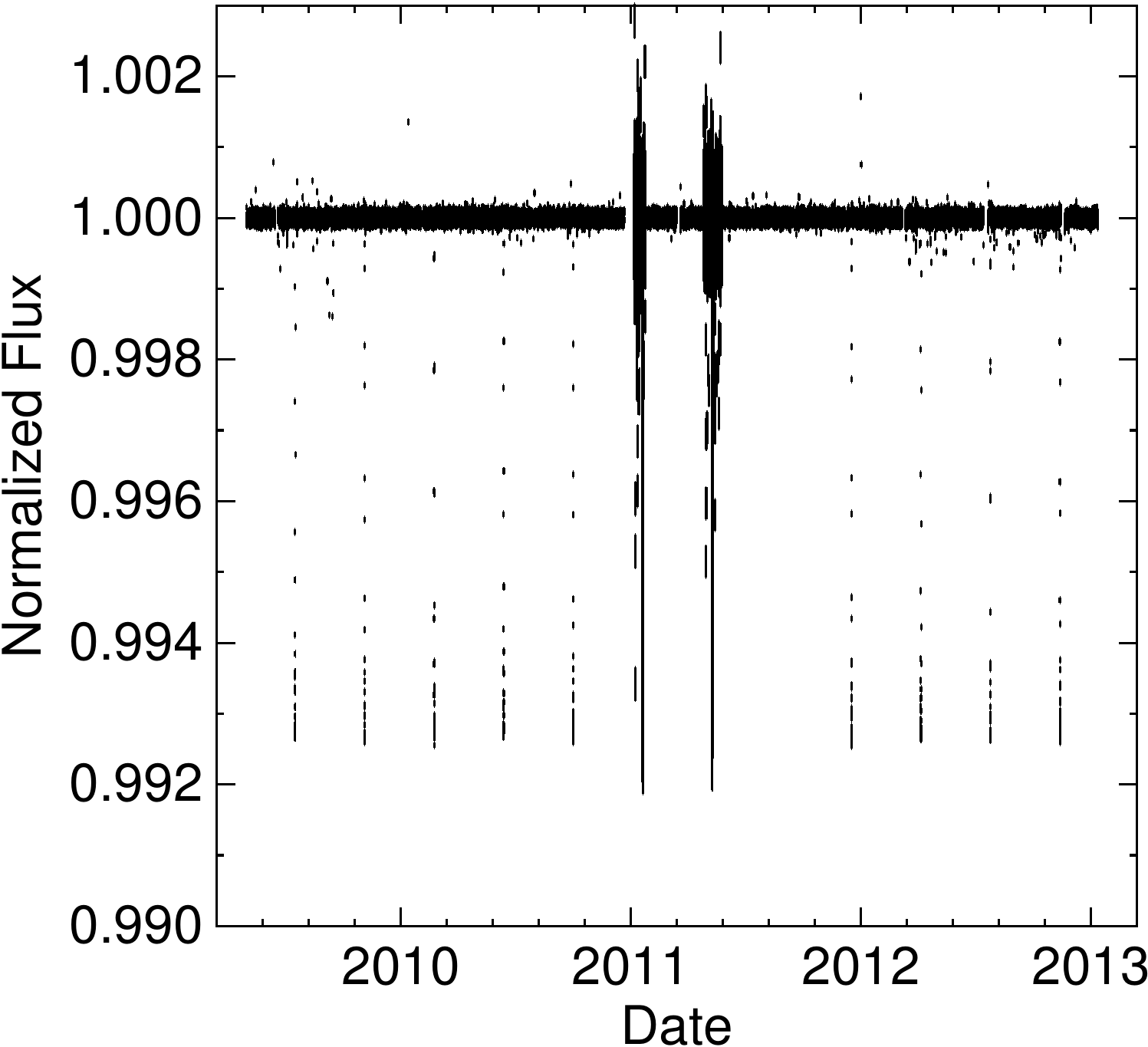}
\caption{\footnotesize This figure shows a boxcar-filtered version of all of the \star~photometry that we use in this paper, combined into a single dataset before period folding. The vertical extent of each data point indicates its error bar. The noisier data in 2011 correspond to short cadence observations, which use a time integration of 1 minute rather than the 30 minutes used in the long cadence observations. The typical photometric precision of the long-cadence data is $4.0\times10^{-5}$, and for the short-cadence data is $2.3\times10^{-4}$.}
\end{figure}

We fold the short and long cadence data sets on their 110-day orbit periods \citep{2011ApJ...736...19B} and crop to a window 26.6 hours long centered on the time of inferior conjunction to arrive at a light curve for fitting. We then average the long cadence data into 30 minute bins and the short cadence data into 5 minute bins, and combine the two data sets into a single light curve. We perform this binning process to increase computational time; we compared an analytical fit of the binned vs. unbinned data to ensure no vital information was lost. We then clean the data to remove any remaining outliers, and begin the fitting process.

\renewcommand{\arraystretch}{1.5}
 \begin{table}[tbhp]
 \centering
 \begin{tabular}{|c|c|}
 \hline
 Parameter & Previously\\ 
 & Reported Value \\ \hline
 $T_0$ & 130.6345 days $\pm0.00015$ \\ \hline
 Period & 110.32160 days $\pm0.00005$\\ \hline
 a & 0.581 AU \\ \hline
 $T_{\mathrm{eq}}$ & 754 K \\ \hline
 Duration & 13.32 h \\ \hline
 Depth & 7270 ppm \\ \hline
 $\frac{\mathrm{d}}{\mathrm{R}_{\star}}$ & 51.19 $\pm 0.14$ \\ \hline
 SNR & 1658.2 \\ \hline
 $T_{\mathrm{eff}}$ & 9257 K \\ \hline
 $m_\mathrm{Kepler}$ & 11.375 \\ \hline
 $\log(g)$ & 4.13 \\ \hline
 $v\sin(i)$\footnotemark[1] & 90 km/s \\ \hline
 \end{tabular}
 \caption{\footnotesize \label{table:parameters}
Transit parameters measured by the Mikulski Archive for Space Telescopes for the \candidate~system. The time of transit center is denoted as $T_0$, the semi major axis of orbit is denoted as a, the equilibrium surface temperature of the planet is denoted as $T_{\mathrm{eq}}$, the duration is the transit duration, the depth is transit depth at center of transit, the ratio of the planet-star separation at the time of transit to the stellar radius is denoted by $d/R_{\star}$, SNR is the signal to noise ratio, the stellar effective temperature is denoted by $T_{\mathrm{eff}}$, and the log of stellar surface gravity is denoted by $\log(g)$. $^a$From the Kepler Community Follow-up Observing Program,  $v\sin(i)$ is the projected stellar rotational velocity.}
\end{table}

\section{\textbf{MODEL}}\label{section:model}

We model the KOI-368 transit using an algorithm developed by \citet{oblateness.2003} and modified to treat rapidly-rotating, oblate stars \citep{2009ApJ...705..683B}. The asymmetry of the non-uniformity in flux coming from a gravity-darkened stellar disk drives the use of explicit numerical integrals to compute eclipsed flux rather than an analytical expression \citep{2002ApJ...580L.171M}.  

Our {\tt transitfitter} program  \citep{oblateness.2003, 2009ApJ...705..683B}  computes this integral for the stellar flux $F$ in polar coordinates $R$ (the projected distance from the center of the star) and $\theta$ (azimuthal angle counterclockwise from right) as

\begin{equation}
F = 1 - \frac{\int_0^{R_{\star}}\int_0^{2\pi}I(r,\theta)~\Gamma(r,\theta) r~d\theta~dr}            {\int_0^{R_{\star}}\int_0^{2\pi}I(r,\theta)~ r~d\theta~dr}
\end{equation}

where $I(r,\theta)$ is the flux per unit area of the particular ($r$, $\theta$) location on the stellar disk, and $\Gamma(r,\theta)$ is a function that denotes the location of the transiting object by yielding 1 for blocked locations and 0 everywhere else. Functionally we evaluate the $\theta$ portion of the integral by finding both limbs of the planet at distance $r$ by use of a root-finding routine and then integrating $I(r,\theta)$ between these known limb locations.

We fit this model to the \emph{Kepler} data using a Levenberg-Marquardt $\chi^2$ minimization algorithm \citep{NumericalRecipes}.  Because of the need for explicit numerical integrals, each fit takes several days to complete.

\section{\textbf{RESULTS}}\label{section:results}
We measured seven different parameters for KOI-368 by fitting its \emph{Kepler} transit light curve. We measured the stellar and planet radii ($R_{\star}$ and $R_\mathrm{p}$, respectively), the inclination of the orbit relative to the plane of the sky, $i$, the time of inferior conjunction, $T_0$, the out-of-transit normalized stellar flux, $F_0$, the sky-projected spin-orbit alignment, $\lambda$, and the stellar obliquity, $\psi$, measured as the tilt of the stellar north pole away from the \emph{Kepler} field of view.

We also derive three parameters based off of our best-fit values. We calculated the impact parameter, $b$, the stellar rotation period, $P_{rot}$, and the stellar oblateness $f_{KOI-368}$. The impact parameter was derived using the orbital inclination angle. The stellar rotation period was derived using our assumed $v$sin$(i)$ value and the stellar obliquity. The stellar oblateness was derived from $R_{\star}$ and $P_{rot}$. We held our limb darkening parameter $c_{1}$ constant at 0.49, as explained in section 5.  We list the best-fit and derived values along with their 1$\sigma$ uncertainties in Table 2.

We display the orbital inclination $i$, sky-projected spin-orbit alignment $\lambda$, and stellar obliquity $\psi$ in Figure 2. With the stellar obliquity and sky-projected spin-orbit alignment constrained, we calculated the spin-orbit alignment using 
\begin{equation}
\cos{(\varphi)}=\sin{(\psi)}\cos{(i)}+\cos{(\psi)}\sin{(i)}\cos{(\lambda)}
\end{equation}
\citep{2011ApJS..197...10B}. We calculated its uncertainty using a Monte Carlo numeric error propagator. 

\begin{figure}[tbhp]
\epsscale{1.1}
\plotone{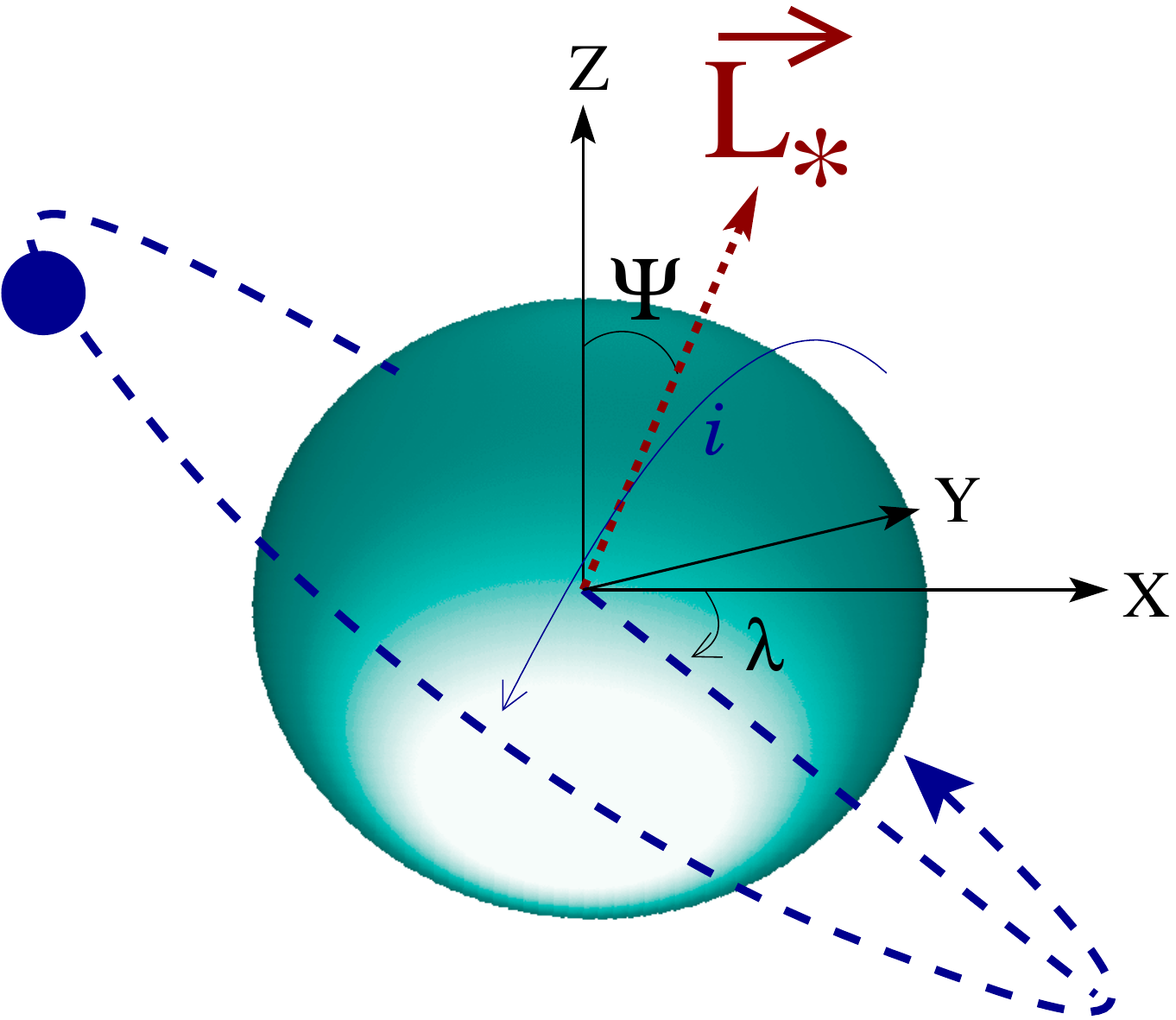}
\caption{\footnotesize Definitions of our angular geometric quantities. The planet's orbital inclination is $i$. The planet's projected spin-orbit angle is $\lambda$, as measured clockwise from stellar east. The coordinate system axes are provided, where $Y$ is the direction of the observer's view. The stellar obliquity, $\psi$, is measured as the angle that the north stellar pole is tilted away from the plane of \emph{Kepler's} view. 
\label{angles and geometry}}
\end{figure}

\renewcommand{\arraystretch}{1.5}
\begin{table}[!htbp]
\centering
\begin{tabular}{|c|c|}
\hline
Parameter & Best Fit Values \\ \hline
$\chi^2_\mathrm{reduced}$ & 1.41 \\ \hline
$R_{\star}$ & $ 2.28\pm.02~\mathrm{R_\odot}$ \\ \hline
$R_\mathrm{p}$ & $1.83\pm0.02~\mathrm{R_{Jup}}$ \\ \hline
$ \frac{R_\mathrm{p}}{R_{\star}}$ & 0.0823 \\ \hline
$i$ & $89.221^\circ\pm0.013^\circ$  \\ \hline
$b$ & 0.697 \\ \hline
$c_1$ & $0.49$ \\ \hline
$T_0$ & $7520550\pm40$~s \\ \hline
$F_0$ & $1.000024\pm5*10^{-6}$~s \\ \hline
$\lambda$ & $10^\circ\pm2^\circ$ \\ \hline
$\psi$ & $3^\circ\pm7^\circ$ \\ \hline
$\varphi$ & $11^\circ\pm3^\circ$  \\ \hline
$P_{rot}$ & $30.73$ hr \\ \hline
$f_{\star}$ & $0.0275$ \\ 
\hline
\end{tabular}
\caption{\footnotesize \label{table:bestfit}
Transit parameters for the KOI-368 system. $R_{p}$ is in units of equitorial Jupiter radii at one bar level. The time at the center of transit, $T_0$, is measured in seconds after BJD~2454900, after \citet{2011ApJ...736...19B}. Throughout the fitting process, the limb darkening coefficient and gravity darkening parameter were held constant at $c_1=0.49$ and $\beta=0.25$ to remove further degeneracies from our model (see Figure 4).}
\end{table}

The fitted data are plotted in Figure 3, along with the residuals. While the light curve appears relatively symmetric, we know that the star is rotating with a $v$sin$(i)=90$ km/s, which implies that there must be some gravity darkening occuring. It does not show in the light curve due to the transit geometry, which we demonstrate in Figure 4. 

\begin{figure}[htbp]
\epsscale{1}
\plotone{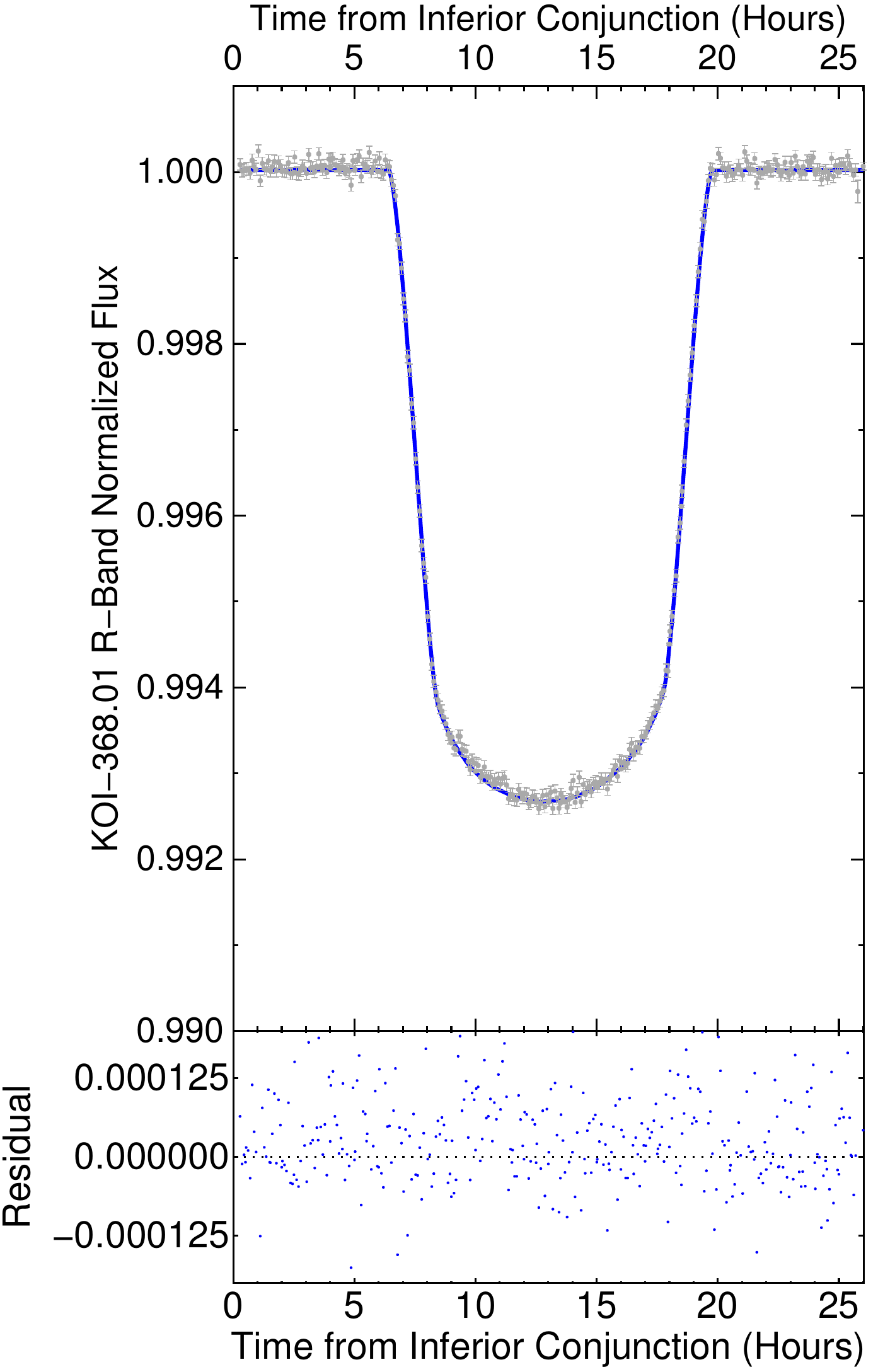}
\caption{\footnotesize Photometry and fits for the 2013 \star~ light curve. We plot the data on top with the gravity-darkened fit in blue. The residuals of this fit are shown below. We recognize a slight asymmetry in the light curve, as first identified by \citet{arXiv:1307.2249v3}. Our gravity-darkened model does a reasonable job of reproducing ingress and egress at the bottom of the light curve. The residuals from the fit are shown at the bottom. The gravity-darkened model does a reasonable job of reproducing the ingress and egress at the bottom of the light curve. \label{fit}}
\end{figure}

\begin{figure}[tbhp]
\epsscale{1}
\plotone{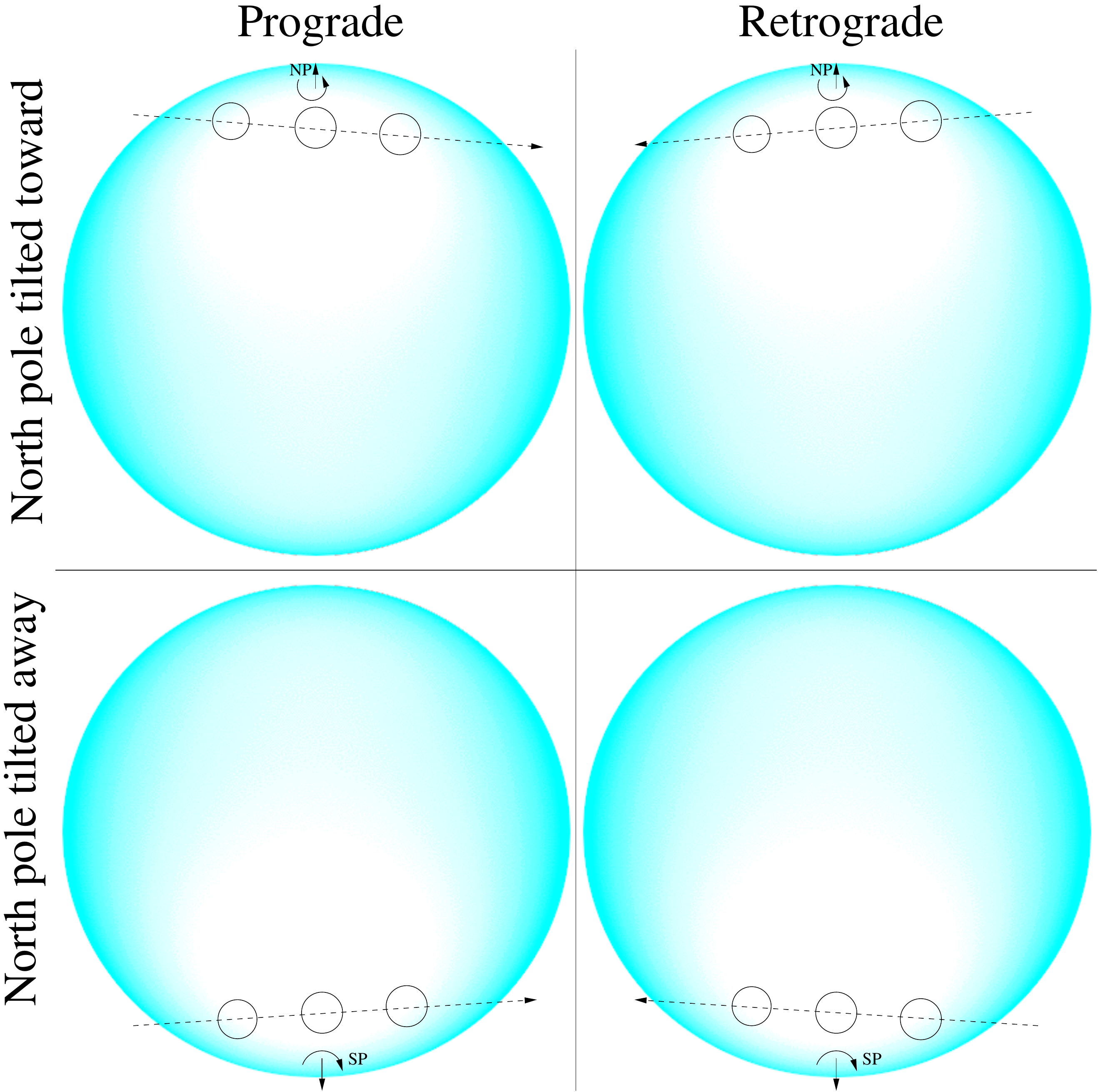}
\caption{\footnotesize The four possible transit geometries of the KOI-368 system. We enhance the effects of gravity darkening and limb darkening to make the varying surface luminosity more evident. All four scenarios produce identical transit light curves; therefore, these geometries are perfectly degenerate. This arises from the inability to differentiate between prograde and retrograde values for the sky-projected spin-orbit alignment $(\lambda$) and the stellar obliquity ($\psi$). Conservatively, we assume prograde values with the north pole tilted toward our point of view (upper left image).  
\label{mergedfigure}}
\end{figure}

\section{\textbf{DISCUSSION}}\label{section:discussion}

\candidate~ has the longest period of any candidate with measured spin-orbit alignment, and is second overall next to \mbox{HD 80606} \citep{arXiv:0906.5605}. Our results show that \candidate~ has the longest orbit that has been proven to be spin-orbit aligned \citep{2012ApJ...757...18A}, as \mbox{HD 80606} is misaligned. This work opens up a whole new population of planets that could be studied via gravity darkening. In particular, we provide a method to determine the possible ways that candidates such as these evolved. It could be that KOI-368 formed in its present location, or it could have migrated inward by some mechanism, while its spin-orbit alignment was left unaffected. This work is the first step in establishing a trend for the evolution and formation of late-type stars, giant planets, and brown dwarfs orbiting hot stars. 

Based on how large the $R_{\star}$ and $R_p$ values are, it is also possible that our fit is assuming that \candidate~ was transiting during apoapsis, thus assuming that the candidate was having to transit across a larger star in order to still hold the $v$sin$(i)$ at a constant of 90 km/s. We can compensate for this by assuming it an eccentric planet. For instance, with an eccentricity of 0.1, we see an $R_{\star}= 2.4864R_{\odot}$ and $R_p=2.0435 R_{Jup}$. When we assume an eccentricity of 0.3, we see an $R_{\star}=3.0633$ and $R_p=2.5176$. Finally, if we were to assume an eccentricity of 0.5, we see an $R_{\star}=3.1374$ and $R_p=2.4363$. However, for our best-fit model, we assumed negligible eccentricity and assumed the transiting companion to be an M-Dwarf star, following \citet{arXiv:1307.2249v3}.

We represented the stellar limb darkening with a single limb darkening parameter, $c_1$, equal to the sum of the two quadratic limb darkening parameters such that $c_1=u_1+u_2$, following \citet{2001ApJ...552..699B}. In our best fit model, we held $c_1$ constant in order to obtain an accurate measure of the stellar obliquity because if $c_1$ were allowed to float, the degeneracy between it and the stellar obliquity would cause the stellar obliquity's uncertainty to greatly increase \citep{2009ApJ...705..683B}. By holding $c_1$ constant, we were able to better constrain the stellar obliquity. While fitting, we used assumed $c_1$ values that varied from 0.43 to 0.56 and found that the best-fit values for the floating parameters varied by less than 1$\sigma$. We found that our reduced $\chi^{2}$ value was lowest at $c_{1}^{2} \approx 0.50$, and thus chose  $c_{1} = 0.49$ based off of the similarity between the stellar radii and temperatures of KOI-386 and KOI-13 \citep{2011ApJS..197...10B}.

We intend to use this technique to survey other systems with intermediate period orbits, and attempt to constrain their spin-orbit alignments. This will allow us to see if \star~ is a good model for these types of systems, or if it is atypical. As future work we will expand the types of systems we analyze, such as smaller radius exoplanets and multiplanetary systems.

\section{\textbf{COMPARISON TO ZHOU \& HUANG (2013)}}
\citet{arXiv:1307.2249v3} recently published constraints for the KOI-368 system using a gravity-darkened model, based off of \citet{2009ApJ...705..683B}. However, \citet{arXiv:1307.2249v3} claims that the KOI-368 system is significantly misaligned ($\varphi$=$69_{-10^{\circ}}^{+9^{\circ}}$), which is contrary to our result that the system is close to alignment. 

This discrepancy may arise from differences between the \citet{arXiv:1307.2249v3} model and ours. Most notably is our handling of the gravity darkening parameter ($\beta$), which relates the effective local gravity to the effective local temperature of a star by \citet{2009pfer.book.....M}:
\begin{equation}
T=T_{pole}\left(\frac{g}{g_{pole}}\right)^{\beta}
\end{equation}

$\beta$ represents the strength of the stellar gravity darkening: higher $\beta$ values allow for larger variations in luminosity between the pole and equator given the same stellar parameters. We use $\beta=0.25$ for our fit, which represents blackbody radiation. \citet{arXiv:1307.2249v3} used a dynamic fit for $\beta$ that arrived at $\beta=0.05$, removing virtually all gravity darkening effects. With such a low $\beta$ value, the stellar obliquity and sky-projected alignment can drastically vary without causing significant asymmetry in a best-fit model. 

The sky-projected alignment, stellar obliquity, limb darkening coefficient, and the gravity darkening parameter are interdependent parameters. \emph{Kepler} photometric data of the KOI-368 system are not sufficiently precise to allow for constraint of all four, so we held $\beta$ constant at 0.25 and $c_{1}$ constant at 0.49 during the fitting process. 

\begin{figure}[tbhp]
\epsscale{1}
\plotone{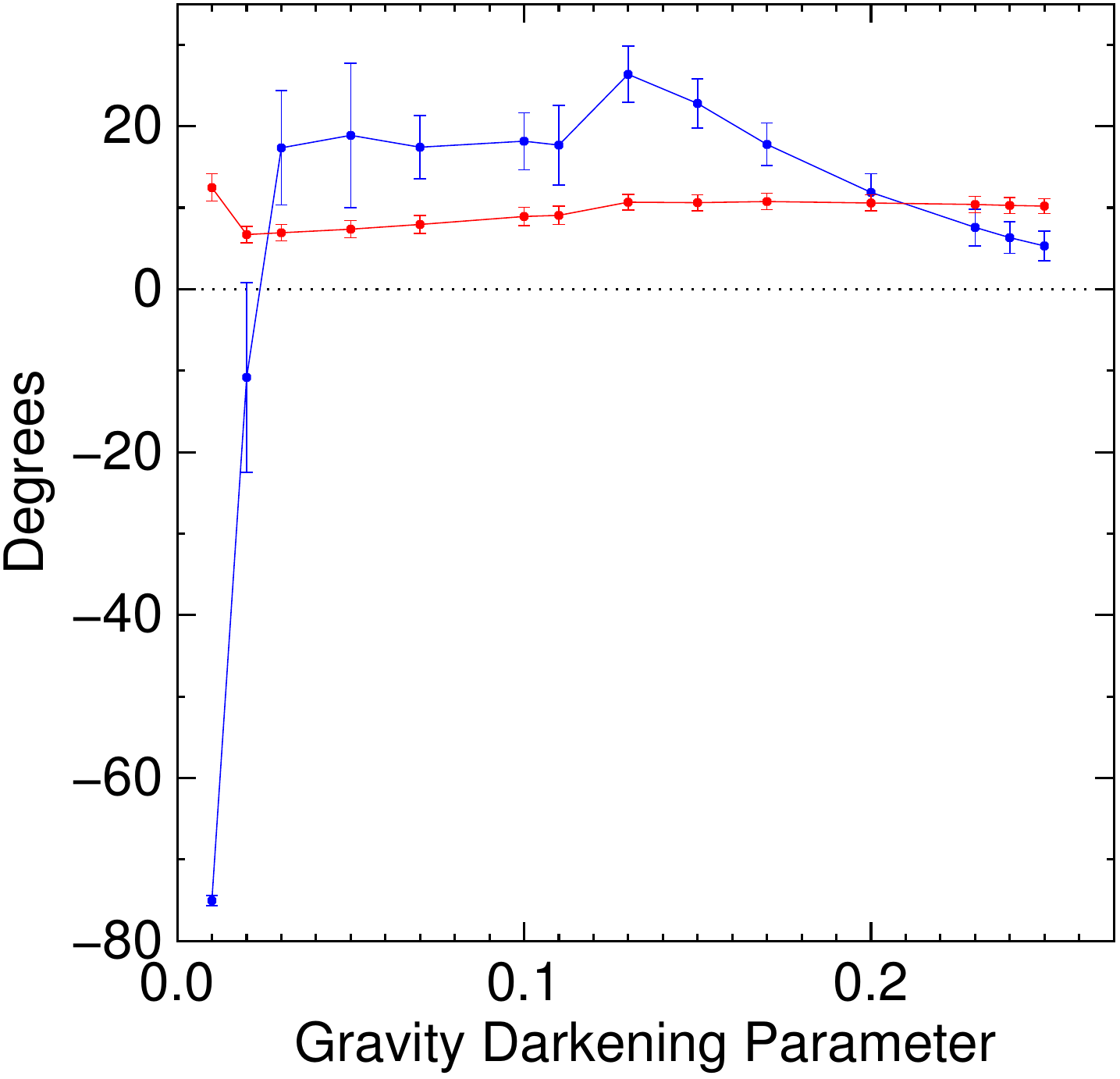}
\caption{\footnotesize This figure shows the $\beta$ dependency of the sky-projected alignment (in red) and of the stellar obliquity (in blue). Doppler tomography of KOI-368  will determine $\beta$, which will allow for better constraint of the spin-orbit alignment. The error bars were determined using constant $\chi^2$ boundaries as confidence limits.
\label{figure:beta}}
\end{figure}

To show the effects of $\beta$ on the fit, we also held $\beta$ constant at other values and constrained $\lambda$ and $\psi$, as shown in Figure 5. This figure shows that the sky-projected alignment and stellar obliquity do not vary extensively except at very low values of $\beta$. Even at $\beta=0.05$, we find a relatively aligned system, which is fundamentally different from \citet{arXiv:1307.2249v3} due to $c_{1}$. 

Our assumption of $\beta=0.25$ is based off of \citet{2011ApJS..197...10B}, which compares the best-fit model of KOI-13 using the theoretical value of  $\beta=0.25$ \citep{1924MNRAS..84..665V} and the experimental value of  $\beta=0.19$ \citep{2007Sci...317..342M}. \citet{2011ApJS..197...10B} found that the best-fit values varied less than $1\sigma$ between the two $\beta$ values. \citet{2011ApJS..197...10B} and Figure 5 suggest that our gravity-darkened model is not significantly varied for small changes in $\beta$. We used the theoretical $\beta$ value, as there currently is no experimental $\beta$ value for the KOI-368 system. A better determination of the star's mass could help to constrain $\beta$ empirically in the future. Various outside measurements would also help to constrain the system, such as a Rossiter-Mclaughlin measurement of the sky-projected alignment, and an asteroseismic determination of the stellar obliquity.

The Zhou \& Huang dynamical fit for both limb darkening and gravity darkening parameter $\beta$ represents an overfit to the \emph{Kepler} data for KOI-368. The resulting fit is unphysical and unreproducible given that Zhou \& Huang do not report the limb darkening parameters that they used. \citet{2009ApJ...705..683B} showed that transits of an aligned planet around a gravity-darkened star lead to anomalies in the best-fit limb darkening coefficients --- particularly the quadratic coefficient $u_2$.

Our fit is grounded in the  \citet{2009ApJ...705..683B} model, which has been shown to to agree with Doppler tomography in the case of KOI-13 \citep{2011ApJS..197...10B, johnson2013doppler}.   \citet{2011ApJS..197...10B} constrained the spin-orbit alignment to $\lambda=24^{\circ}\pm4^{\circ}$, and \citet{johnson2013doppler} constrained it to be $\lambda=21.3^{\circ}\pm0.2^{\circ}$. With an outside confirmation of our model, we think that our measurement of KOI-368 is robust. Future Doppler tomography of KOI-368 could confirm our result. 

\section{\textbf{CONCLUSION}}\label{section:conclusion}
By fitting all available short and long cadence \emph{Kepler} photometry for KOI-368.01, we measured a sky-projected spin-orbit alignment of $\lambda=10^{\circ}\pm2^{\circ}$ and a stellar obliquity of $\psi=3^{\circ}\pm7^{\circ}$. While the limb darkening parameter is assumed to be $c_1=0.49$ \citep{2011ApJS..197...10B}, other fits using different assumed $c_{1}$ limb darkening values while holding $\beta$ constant show that the spin-orbit angle, $\varphi=11^{\circ}\pm3^{\circ}$, is not substantially affected by plausible limb-darkening variations.

The gravity-darkened model allows for determination of the true spin-orbit alignment of a system, not just its sky-projected spin-orbit alignment. This work presents one of the first extrasolar systems to have its spin-orbit alignment constrained \citep{2007AJ....133.1828W, 2011ApJ...743...61S, 2011ApJ...740L..10N}. The spin-orbit alignment of KOI-368.01 does not suggest that bodies orbiting more massive stars are more likely to be spin-orbit misaligned, contrary to \citep{2010ApJ...718L.145W}.  

We show that \candidate~ is well aligned with a spin-orbit alignment  $\varphi=11^{\circ}\pm3^{\circ}$. \star~ is a rapidly rotating star, and therefore the light curve displays the effects of gravity darkening. However, because our system is well aligned, \candidate~ transits across lines of equal brightness; therefore, the light curve displays only nominal asymmetry. 

This system could have formed via one of several mechanisms. The most likely is fragmentation, in which the protostellar disc fragments due to rotational instabilities. This mechanism allows for spin-orbit aligned binary systems of less than 1 AU \citep{arXiv:9411081}. The formation of close-in binary systems is still somewhat unexplained \citep{arXiv:9411081,ApJ-556-265}; the ability to constrain the spin-orbit alignment of such systems will contribute to understanding them. 

The unique nature of the KOI-368 system allows for new insight in studying photometric light curves. With the high precision of \emph{Kepler} photometry, we are for the first time able to constrain systems such as these, which provides new understanding of the formation of extrasolar systems. The knowledge we gained from this system will be applicable to a wide variety of transiting objects in the future.   

\bibliographystyle{apj}
\bibliography{paper}

\end{document}